# PARKER SOLAR PROBE IN-SITU OBSERVATIONS OF MAGNETIC RECONNECTION EXHAUSTS DURING ENCOUNTER 1

T. D. Phan[1], S. D. Bale[1,2,3], J. P. Eastwood[3], B. Lavraud[4], J. F. Drake[5], M. Oieroset[1], M. A. Shay[6], M. Pulupa[1], M. Stevens[7], R. J. MacDowall[8], A. W. Case[7], D. Larson[1], J. Kasper[9], P. Whittlesey[1], A. Szabo[8], K. E. Korreck[7], J. W. Bonnell[1], T. Dudok de Wit[10], K. Goetz[11], P. R. Harvey[1], T. S. Horbury[3], R. Livi[1], D. Malaspina[12], K. Paulson[7], N. E. Raouafi[13], and M. Velli[14]

[1] SSL, University of California, Berkeley, CA 94720, USA; phan@ssl.berkeley.edu
[2] Physics Department, University of California, Berkeley, CA 94720-7300, USA
[3] The Blackett Laboratory, Imperial College London, London, UK
[4] IRAP, Université de Toulouse, France
[5] University of Maryland, College Park, MD, USA
[6] University of Delaware, Newark, DE, USA
[7] Smithsonian Astrophysical Observatory, Cambridge, MA, USA
[8] NASA Goddard Space Flight Center, Greenbelt, MD, USA
[9] Climate and Space Sciences and Engineering, University of Michigan, Ann Arbor, MI, USA
[10] LPC2E, CNRS and University of Orléans, Orléans, France
[11] School of Physics and Astronomy, University of Minnesota, Minneapolis, MN, USA
[12] University of Colorado LASP, Boulder, Colorado, USA
[13] Johns Hopkins University Applied Physics Laboratory, Laurel, MD, USA
[14] University of California, Los Angeles, Los Angeles, CA, US

Short Title: Magnetic Reconnection during PSP 1st Encounter

Corresponding Author: Tai Phan, phan@ssl.berkeley.edu

**Abstract**

Magnetic reconnection in current sheets converts magnetic energy into particle energy. The process may play an important role in the acceleration and heating of the solar wind close to the Sun. Observations from Parker Solar Probe provide a new opportunity to study this problem, as it measures the solar wind at unprecedented close distances to the Sun. During the 1$^{st}$ orbit, PSP encountered a large number of current sheets in the solar wind through perihelion at 35.7 solar radii. We performed a comprehensive survey of these current sheets and found evidence for 21 reconnection exhausts. These exhausts were observed in heliospheric current sheets, coronal mass ejections, and regular solar wind. However, we find that the majority of current sheets encountered around perihelion, where the magnetic field was strongest and plasma β was lowest, were Alfvénic structures associated with bursty radial jets and these current sheets did not appear to be undergoing local reconnection. We examined conditions around current sheets to address why some current sheets reconnected, while others did not. A key difference appears to be the degree of plasma velocity shear across the current sheets: The median velocity shear for the 21 reconnection exhausts was 24% of the Alfvén velocity shear, whereas the median shear across 43 Alfvénic current sheets examined was 71% of the Alfvén velocity shear. This finding could suggest that large, albeit sub-Alfvénic, velocity shears suppress reconnection. An alternative interpretation is that the Alfvénic current sheets are isolated rotational discontinuities which do not undergo local reconnection.

1. **INTRODUCTION**

Magnetic reconnection in current sheets is a universal plasma process that converts magnetic energy into plasma jetting and heating, and is important in many laboratory, space, solar, and astrophysical contexts (e.g., Yamada 2010; Paschmann et al. 2013; Parker 1983; Priest 1984; Duncan and Thompson 1992; Hurley et al. 2005; Kronberg 2002; Abdo 2011). For example, reconnection plays an important role transferring solar wind mass, momentum and energy into the magnetosphere of Earth (e.g., Dungey 1961) and other planets (e.g., Slavin and Holzer 1979). On the Sun, it has been suggested that reconnection plays an important role in the heating of the corona (e.g., Parker 1988; Klimchuk 2006), in the initiation of solar flares (e.g., Giovanelli, 1946; Parker 1963; Antiochos 1999), as well as the restructuration of the corona over the solar cycle (e.g., Babcock, 1961; Leighton, 1969; Owens et al., 2007; Lavraud et al., 2011). In the solar wind, in-situ observations by spacecraft at 1AU and beyond have revealed the presence of reconnection exhausts in current sheets in the vicinity of interplanetary coronal mass ejections (ICME) (e.g., Gosling et al. 2007a; Gosling and Szabo 2008; Phan et al. 2009; Ruffenach et al. 2012, 2015; Lavraud et al. 2014), at heliospheric current sheets (e.g., Gosling et al. 2005b, 2006a; Lavraud et al. 2009), and in the regular solar wind (e.g., Gosling et al., 2005a, 2006b; Gosling 2007; Phan et al. 2006; 2010; Davis et al. 2006; Eriksson et al. 2009, Mistry et al. 2015; 2017).

At 1 AU, reconnection is detected in only a small fraction of solar wind current sheets encountered by spacecraft; thus reconnection is not energetically important at 1AU in terms of the evolution of Heliospheric plasmas and fields (Gosling 2005a, 2007a). However, it is predicted that the occurrence rate of reconnection in solar wind current sheets could be much higher closer to the Sun because of the lower plasma β environment there, which lessens constraints on the occurrence of reconnection associated with the magnetic geometry of the current sheet (Swisdak et al. 2003; 2010; Phan et al. 2010).

Parker Solar Probe (PSP) (Fox et al. 2016) provides an unprecedented opportunity to probe the physics and properties of solar wind reconnection in the near-Sun environment. During the 1$^{st}$ orbit, PSP's closest approach was 35.7 solar radii ($R_s$), allowing reconnection to be explored at such close distances to the Sun for the first time. We have performed a comprehensive survey of plasma and field measurements to determine the occurrence of reconnection exhausts, providing a first assessment of the role reconnection might play in the solar wind close to the Sun. We present examples of reconnection exhausts in and around an Interplanetary Coronal Mass Ejection (ICME), Heliospheric Current Sheets (HCS) crossings, and the regular solar wind. We find no evidence of reconnection near Orbit 1 perihelion; at this time the solar wind appeared to be dominated by Alfvénic structures, rather than reconnection in the local current sheets. Whether the absence of reconnection near perihelion in this orbit is due to the particular type of solar wind encountered, or is intrinsic to the near-Sun environment, remains to be understood.

The paper is organized as follows: Section 2 describes the dataset and methodology used. In Section 3 we show an overview of Encounter 1. In Section 4 examples of reconnection exhausts in different environments are presented. We find that the sub-second resolution plasma and field measurements are often required to resolve thin current sheets encountered during Orbit 1. Section 5 shows that plasma jets seen near perihelion are mostly associated with Alfvénic structures, rather than reconnection in the local current sheets. Section 6 discusses statistical properties of current sheets associated with reconnection versus Alfvénic structures. Section 7 summarizes and discusses our findings.

## 2. INSTRUMENTATION, EVENT SELECTION CRITERIA, AND CURRENT SHEET COORDINATE SYSTEM

### 2.1. Instrumentation

We have examined PSP data from 2018-10-27 to 2018-11-25, covering radial Solar distances of 75.2 $R_s$ to 35.7 $R_s$ on the inbound, and to 111.8 $R_s$ on the outbound legs of Orbit 1 (Figure 1). Magnetic field data is measured by the FIELDS fluxgate magnetometer (Bale et al., 2016). The sampling rate of magnetic field data was 290 vectors/s near perihelion, and gradually lower at larger radial distances, with cadence of 0.2 s at 111 $R_s$. For uniformity of data presentation and for the comparison of current density values at various radial distances (Figure 1e), the magnetic field data presented in the present paper has been averaged to 0.2 s.

Proton velocity and density used in this paper are measured by the SWEAP/SPC instrument (Kasper et al., 2016; Case et al., 2019). Here we use the best resolution data available, which was 0.22s cadence just prior to the central perihelion pass (2018-11-01/15:28 to 2018-11-03/11:51 UT), 0.9s in the intervals either side (2018-10-31/03:00 to 2018-11-01/15:28 UT before, 2018-11-03/11:51 to 2018-11-11/03:28 UT after). Outside of these times, the SPC proton moment cadence was ~28s. Electron pitch angle fluxes are measured by the SWEAP/SPAN instrument (Whittlesey et al., 2019). The SWEAP/SPAN electron pitch angle data we use have a cadence of 28s from 2018-10-31/03:00 UT to 2018-11-12/03:30 UT. The cadence was 900 s outside this interval.

### 2.2. Reconnection and non-reconnection event identification

We identify reconnection exhausts in current sheets by the presence of accelerated plasma flows. These flows are bounded on one edge of the current sheet by correlated changes in velocity, **V**, and magnetic field, **B**, and anti-correlated changes in **V** and **B** on the other edge, indicating a pair of rotational discontinuities (RDs) bounding the exhaust. Such opposite correlations between the changes in **V** and **B** are consistent with Alfvénic disturbances propagating in opposite directions along reconnected field lines away from the reconnection site (X-line) (e.g., Gosling et al. 2005a). We do not identify reconnection events based on magnetic field profiles such as deep minima in |**B**| or bifurcated current sheets alone, as they are not unique to reconnection events. This means that events can only be identified if their duration is longer than the cadence of the plasma measurements. Examples of reconnection events illustrating the method are shown in Figures 2-4.

In contrast, non-reconnecting current sheets are recognized by the absence of plasma jetting within the current sheets. In such cases, the current sheet could be a tangential discontinuity (TD) or a single rotational discontinuity (RD). Across a TD, the **V** and **B** variations are usually not correlated since the plasmas on the two sides of a TD are not magnetically connected with each other. Across a single RD, there should be a single correlation between **V** and **B.** Examples of non-reconnecting current sheets are shown in Figure 5.

### *2.3. Current sheet (LMN) coordinate system*

In the present paper, the data are shown in two coordinate systems. The large-scale context data are shown in the spacecraft centered RTN coordinate system (Hapgood, 1992), where **R** is the direction from the Sun to the spacecraft, **T** is the cross product of the Sun's rotation vector with **R**, and **N** = **R** × **T**. Individual current sheets are shown in a local current sheet (XYZ) coordinate system. This uses a hybrid minimum variance method, which has been found to work well in low magnetic shear (large guide field) current sheets (Gosling and Phan 2013). The current sheet normal direction, **Z**, is determined from $\mathbf{B_1} \times \mathbf{B_2} / |\mathbf{B_1} \times \mathbf{B_2}|$, where **B₁** and **B₂** are the magnetic field vectors at the two edges of the current sheet. **Y** = **Z** × **X'** is approximately the out-of-plane X-line direction, where **X'** is the maximum variance direction from the minimum variance of the magnetic field (MVAB) analysis (Sonnerup and Cahill 1967), **X** = **Y** × **Z** is approximately along the anti-parallel magnetic field direction.

### 3. OVERVIEW OF ENCOUNTER 1

Figure 1 shows the large-scale context of the PSP in-situ observations of the solar wind from 2018-10-27 to 2018-11-25, with magnetic field magnitude (Panel a) increasing with decreasing distance to the Sun, reaching ~ 100 nT near perihelion (see Bale et al. 2019 for details on the radial dependence of the field magnitude). Large fluctuations in the radial component of the magnetic field (Panel b) are seen throughout the encounter. Figure 1e reveals these fluctuations to be associated with current density spikes, reaching $|\mathbf{j}| \sim 0.6$ µA/m$^2$ near perihelion on November 5-8. Such peak current densities across magnetic field variations δB up to 100 nT imply current sheet widths (w ~ ΔB/µ₀j) of the order of 100 km, compared to the ion inertial length of ~15 km in this interval. Closer inspection of the current density and magnetic field variations throughout the interval in Figures 1 reveals a range of current sheet thicknesses, from kinetic to MHD scales.

Reconnection exhaust jet speed is expected to scale as the inflow Alfvén speed, and the ambient solar wind Alfvén speed reached an average level of ~ 150 km/s near perihelion at 35.7 $R_S$ (around 2019-11-06), more than a factor of three larger than at 110 $R_s$ (Figure 1d). This indicates that reconnection exhausts, if present, should be easily identifiable in the data.

To set the context for the reconnection exhausts and Alfvénic structures to be described in the next section, certain key regions in which reconnection was detected are labeled in Figures 1a-c. HCS crossings are recognized by polarity reversals of the large-scale radial magnetic field (Figure 1b) (Bale et al. 2019; Szabo et al. 2019), together with the concurrent switching between $0^o$ and $180^o$ pitch angle of fluxes of 320 eV electrons, which are strahl electrons of solar origin. Two ICMEs encountered by PSP during Orbit 1 (Korreck et al. 2019; Nieves-Chinchil et al. 2019) are also marked in Figure 1a.

## 4. OBSERVATIONS OF RECONNECTION EXHAUSTS

We examined the entire 29-day interval shown in Figure 1 to search for evidence for reconnection exhausts. Table 1 lists all 21 reconnection exhausts that we have been able to identify so far in Orbit 1, and their properties in terms of exhaust width, magnetic and velocity shear values, and the locations and regions in which they were detected. We now show some of the cleanest examples of exhausts detected by PSP, organized by the event environments.

### *4.1. Reconnection in interplanetary coronal mass ejections*

There were two prominent ICMEs detected by PSP during Orbit 1 (Figure 1a). The October 31 ICME is smaller than that on November 11-12 (in terms of spatial dimension and magnetic field strength enhancement), but it has the advantage that high time-resolution (0.9s) proton moment data are available.

Figures 2a and b show the large-scale structure of the October 31 ICME, detected at ~55 $R_s$ as the structure moved past PSP, with a large enhancement in the magnetic field magnitude |**B**| up to 80 nT. Large rotations of the magnetic field are seen inside the ICME between 08:15 and 08:30 UT on Oct 31, 2018, although small changes in the magnetic field exist throughout. We found three occurrences of reconnection in current sheets within the ICME. Interestingly, only one of these occurred in a large rotation of the magnetic field, while the other two had extremely small magnetic field rotations. We now describe the 3 events in detail. The data are presented in the XYZ current sheet coordinate system (see Section 2).

Figure 2c-g shows an event where PSP detected the passage of current sheet (panel d), with embedded reconnection outflow at 06:43:21-06:43:27 UT (between the two vertical dashed lines). The current sheet crossing duration was only 5.7s. The exhaust was identified by the presence of accelerated flow in the positive X direction (panel f) within the region where the magnetic field rotated (panel d), with the change in proton velocity in the outflow direction $V_{pX}$ (panel f) and the reconnecting field component $B_X$ (panel e) being correlated on the leading edge and anti-correlated on the trailing edge of the exhaust, as described in Section 2. The observed changes in the outflow velocity $V_{px}$ were ~24 km/s and ~12 km/s across the leading and trailing edges of the exhaust (going from outside to the middle of the current sheet), respectively, and these changes were ~64% and 59% of the predicted changes at the corresponding edges of the exhaust according to the rotational discontinuity jump condition (Hudson 1970):

$$V_{pX2} - V_{pX1} \sim \pm (B_{X2} - B_{X1})/(\mu_0 \rho_1)^{1/2} \tag{1}$$

where $V_{pX}$ and $B_X$ are the X component of the proton bulk velocity and magnetic field, and $\rho$ is proton mass density (here we omit the pressure anisotropy effect for simplicity). Subscripts 1 and 2 denote the inflow and outflow regions, respectively. The positive and negative signs of this relation refer to the leading and trailing edges of the exhaust, respectively, for this example. The sub-Alfvénic outflow is a common feature of reconnection in observations (e.g., Sonnerup et al., 1981; Paschmann et al., 1986; Øieroset et al., 2000) and in kinetic simulations (e.g., Liu et al., 2012; Haggerty et al., 2018), and has been attributed to the fact that the exhaust boundaries are not pure rotational discontinuities (RDs), but a combination of RDs and slow shocks (e.g., Lin and Lee, 1993; Teh et al., 2009; Liu et al., 2012). It could also be due to the effect of ion temperature anisotropy in the exhaust (Haggerty et al., 2018).

In this event, the reconnection outflow detected by PSP was radially anti-sunward, leading to an enhancement of the radial flow speed relative to the ambient solar wind flow (Figure 2g). This implies that the X-line was located sunward of PSP.

The magnetic shear angle ($\theta$) across the current sheet was only ~ 37°, i.e., the guide field $B_Y$ was 3.2 times the reconnecting field $B_X$. A characteristic of strong guide field current sheets is that |**B**| is typically only very slightly depressed inside the reconnecting current sheet, as observed here (see also Gosling & Szabo 2008; Phan et al., 2010) because the guide field is modestly compressed within the exhaust to maintain total pressure balance (Lin and Lee 1993; Zhang et al., 2019). The sharp changes in the magnetic field orientation near the two edges and a plateau in between indicate that the current sheet was bifurcated (e.g., Phan et al. 2006; Gosling & Szabo 2008). The $B_X$ level of the plateau being above halfway point between the two asymptotic $B_X$ is due to the presence of tangential $V_{pX}$ velocity shear (Eriksson et al. 2009), such that the jumps in $V_{pX}$ and $B_X$ are larger at the leading edge of the exhaust.

Finally, it is noted that the shear of the velocity component along the reconnecting field direction, $\Delta V_{pX}$, measured by the difference between $V_{pX}$ at the two dashed lines (Figure 2f), was ~10.6 km/s. This is 18% of the shear in the Alfvén velocity (58 km/s) based on the reconnecting field component $B_X$. The velocity shear of the transverse (to reconnecting field) velocity $\Delta V_{pY}$ was much smaller (2.5 km/s) in this case.

The second example of a reconnection exhaust in the Oct 31 ICME is shown in Figures 2h-l. This event has an even larger guide field $B_Y$ (4.2 times the reconnecting field $B_X$), i.e., the magnetic field rotation angle was only 29º. Because the magnetic shear is so small, the event is nearly unrecognizable on the scale of Figure 2b and the magnetic field magnitude is nearly unchanged across the current sheet. The reconnection exhaust is recognized by the proton jetting in the negative X direction (panel k), with $B_X$ and $V_{pX}$ being correlated across the left edge, and anti-correlated across the right edge of the current sheet. The reduction in the proton radial velocity (panel l) inside the current sheet implies that here the X-line was located anti-sunward of PSP. The $V_{pX}$ velocity shear (or difference) on the two sides of the current sheet again leads to an asymmetry in the amount of velocity change across the two edges of the current sheet.

The third example (Figure 2m-q) had larger magnetic shear than the other previous two events and the large rotation of $B_X$ stands out in Figure 2b. The |**B**| reduction inside the current sheet is more visible in this event but is mitigated by the compression of $B_Y$. However, the magnetic shear was still only 90º, i.e., a guide field of unity. The duration of the current sheet crossing was 7.6 s. This current sheet was located near the compressed trailing edge of the ICME (Figure 2a). The reason for the compression appears to be due to the fact the ICME was moving slower (radially) than the solar wind behind it, a phenomenon that is known to occur frequently in ICMEs at 1 AU (e.g., Fenrich and Luhmann 1998; Ruffenach et al., 2015). The event is again characterized by a bifurcated current sheet (sharp changes in $B_X$ at the 2 edges). The out-of-plane magnetic field $B_Y$ was enhanced inside the current sheet, with small dips at the two edges. Such tripolar $B_Y$ profiles have previously been reported in some solar wind reconnection events (Eriksson et al. 2014, 2015) and in simulations (Eriksson et al. 2015; Zhang et al., 2019). A velocity shear across the current sheet was also present in this event, which again leads to the plateau in $B_X$ being below the half point of $B_X$ transition. Finally, the reduction in $V_{pX}$ in the current sheet indicates an anti-sunward location of X-line relative to PSP.

### *4.2. Reconnection at Heliospheric Current Sheets*

Figures 3a and b show that PSP crossed the HCS a number of times during the first orbit, marked by reversals in the $B_R$ component (panel a) and the concurrent switching between 0° and 180° pitch angle of fluxes of 320 eV strahl electrons (panel b). All of these occurred in regions where SPC had 28 s resolution proton moments, thus not all current sheets could be resolved by the proton measurements. Nevertheless, we have found possible evidence for reconnection exhausts in the HCS on November 13-15, and on November 23.

The HCS on November 13-15 seems to be composed of multiple current sheets. Figures 3c-q display examples of reconnection exhausts in three of the current sheets. All 3 events had relatively high magnetic shears: 129° (panel c-g), 168° (panel h-l) and 108° (panel m-q). The 168° shear one is by far the widest of the three current sheets (see Table 1). It also had a deep |**B**| minimum (Figure 3h), whereas the 129° and 108° events did not. All 3 events show the presence of accelerated flows consistent with reconnection: opposite $\delta V_{pL}$-$\delta B_L$ correlations at the 2 edges of the current sheet.

In contrast, the November 23 event (Figures 3r-v) was a single crossing of the HCS. The crossing occurred at ~107 $R_s$ from the Sun. The magnetic field rotation across the current sheet was 168°, i.e. the magnetic fields on the two sides of the current sheet were nearly anti-parallel. There was a deep minimum in |**B**| in the current sheet (Figure 3r), which is a characteristic of high-magnetic shear current sheets. Embedded in the current sheet is a proton jet of ~ 40 km/s in the X direction (relative to the external flows), with opposite $\Delta V_{pX}$-$\Delta B_X$ correlations at the 2 edges of the current sheet, consistent with reconnection. The enhancement of the radial velocity $V_{pR}$ implies that the X-line was located sunward of PSP. The duration of the HCS crossing was ~11 minutes, which translates to an exhaust width of $1.7 \times 10^5$ km, or ~6000 ion inertial lengths.

Interestingly, in these 4 HCS events, the radial velocity $V_{pR}$ was enhanced as a result of reconnection outflows, which implies that the X-line was sunward of PSP in all cases.

### 4.3. Reconnection in regular solar wind

In addition to reconnection exhausts occurring in ICMEs and HCS crossings, PSP also detected reconnecting current sheets in the regular solar wind. Figures 4 c-l show two such examples. One event (Figure 4h-l) was detected when the ion moment resolution was 0.22 s and the cleanly-resolved plasma jet illustrates the quality of this high-resolution data product.

The current sheets in both events were bifurcated (Panels e,j), and plasma jetting was observed in the current sheet (panels f,k), with opposite $\delta V_{pX}$-$\delta B_X$ correlations at the 2 edges of the current sheets (panels e,f and j,k). Again, because of the relatively large guide field, |**B**| did not have a deep minimum in these reconnecting current sheets.

### 4.4. Summary of reconnection exhaust observations

The above examples and additional examples in Table 1 illustrate that reconnection exhausts occur in a variety of solar wind phenomena that PSP encountered. They also illustrate the various conditions (e.g., magnetic and velocity shears) under which reconnection occurs. Finally, the plasma and field profiles across reconnecting current sheets are dependent on the guide field and velocity shear. Here, we summarize some key observations:

- There is a large range of magnetic shear associated with reconnecting current sheets detected by PSP, ranging from 27° to 168°, with a median of 89°. The largest shears were at HCSs.

- Only the largest magnetic shear events displayed deep minima in |**B**|.

- The majority of reconnecting current sheets are bifurcated. However, as will be shown in the next section, bifurcated current sheets do not necessarily imply reconnection.

- The presence of velocity shear across the current sheet $\Delta V_{pX}$ (in the flow component parallel to the reconnecting field) creates an asymmetry in the level of plasma acceleration at the two edges of the current sheet.

- Approximately half of the reconnection events resulted in enhanced radial velocity $V_{pR}$, while the others showed reduced $V_{pR}$. This is consistent with reconnection occurring in the local current sheet, such that statistically there is an equal chance for PSP to be on either side of the X-line.

- The distance from PSP to the reconnection site along the outflow direction (X) can be roughly estimated as $(W/2)/R_{rec}$, where W is the width of the exhaust (along Z) at PSP, and $R_{rec}$ is the dimensionless reconnection rate. Table 1 shows the estimated distance using the canonical reconnection rate of 0.1 (e.g., Birn et al. 2001). The distances to the X-line were large for the two main HCSs (up to 2 $R_s$), but much smaller for other events.

- Many reconnecting current sheets were thin, (with crossing duration << 1 minute). Adequately resolving such current sheets requires high-time (sub-second) resolution plasma and field measurements.

## 5. NON-RECONNECTING CURRENT SHEETS ASSOCIATED WITH RADIAL JETS AND RADIAL MAGNETIC FIELD SWITCHBACKS NEAR PERIHELION

Table 1 lists reconnection events that we have been able to identify in Encounter 1 data so far. It is noted that there are no events found in the interval between 2018-11-02/13:05:07 and 2018-11-11/23:17:00 UT, which encompasses perihelion. This is surprising because this is the interval with the strongest magnetic fields (Figure 1a), largest Alfvén speeds (Figure 1d), and current sheets with the highest levels of current density (Figure 1e) of Orbit 1. Furthermore, high resolution (≤ 0.9 s) plasma and (<< 0.2 s) magnetic field data were available throughout the interval, thus the lack of reconnection events found cannot be attributed to measurement limitations. This interval is dominated by the presence of radial magnetic field polarity changes (termed 'switchbacks') (Bale et al. 2019) and spiky plasma jets (Kasper et al. 2019; Horbury et al. 2019). We will focus on a representative interval to illustrate the structures of the plasma and magnetic field, and the method used to deduce the absence of reconnection in the local current sheets.

Figures 5a-c show the radial field and the proton radial velocity in a 16-hour interval. It appears that the radial velocity $V_{pR}$ had a baseline velocity of ~ 300 km/s, from which the velocity spiked upward, with enhancements reaching >200 km/s from the baseline. The presence of such jets in a region full of current sheets at first suggests that they could be due to reconnection in the local current sheets. However, the simple fact that the jets were always radially enhanced immediately calls this hypothesis into question. If the jets were due to local reconnection, statistically one would expect roughly half to be radially decelerated since there would be an equal chance of PSP being on either side of the X-line, as is the case for the reconnection events in Table 1.

Another way to determine whether the radial jets are associated with local reconnection or not is to examine the correlation between the variations of $B_R$ and $V_{pR}$ (see Section 2). Figures 5a and b show that $B_R$ and $V_{pR}$ appear to be positively correlated throughout the 16-hour interval. To assess the degree of Alfvenicity of the velocity variations more precisely, Figure 5c and 5d show the overlaid of the observed flows (in black) and the predicted radial velocity if it were Alfvénic (in red). The predicted velocity is computed based on a single reference time (2018-11-06/16:00

UT) marked by the vertical dashed line in panels a-c. The flow prediction is based on the local magnetic field measurements and the reference velocity and field values: $V_{AR,predicted} - V_{pR,ref} = +[B_R (\mu\rho)^{-1/2} - B_{R,ref} (\mu\rho_{ref})^{-1/2}]$ (Hudson 1970; Paschmann et al. 1986). The positive sign was chosen based on the observed positive correlation between the variations of $V_{pR}$ and $B_R$ in this interval. Subscript 'ref' denotes the reference time. The pressure anisotropy effects are omitted for simplicity in the Alfvénic flow prediction.

The agreement between the Alfvén velocity and observed flows is remarkable (Figures 5c and d), even for flow variations that are hours away from the reference time. The single (positive) correlation between the variations of $V_{AR}$ and $V_{pR}$ throughout the 16-hour interval indicates that the $V_{pR}$ enhancements (from a baseline of ~300 km/s) are related to Alfvénic structures, not locally generated reconnection jets bounded by pairs of Aflvénic discontinuities (or RDs) propagating in opposite directions along folded magnetic field lines. The positive correlation persisted for days in regions where the large-scale radial field $B_R$ was negative. Interestingly, the single correlation between $V_{AR}$ and $V_{pR}$ suggests that the plasmas in this entire interval were magnetically interconnected, i.e., there were no topological boundaries between them.

The $V_{AR}$ - $V_{pR}$ correlation becomes negative when the large-scale $B_R$ is positive (not shown). Thus the current sheets near perihelion are Alfvénic structures propagating away from the Sun (Kasper et al. 2019; Bale et al. 2019; Horbury et al. 2019).

To examine the current sheets in Alfvénic structures in more details, Figures 5e-s shows three examples of individual current sheets, viewed in the current sheet XYZ coordinate system. In all three examples, the current sheet was bifurcated, with sharp changes in $B_X$ at the two edges of the current sheet. The guide field $|B_Y|$ was enhanced in each of the current sheets. These are common magnetic field profiles in reconnecting current sheets (see Figures 2-4). However, there was no reconnection plasma jetting inside these current sheet: $V_{pX}$ and $B_X$ variations were positively correlated at both edges, and all the way through each of the current sheets. Such correlations are predicted for a single RD (as opposed to a reconnection back-to-back RDs, see section 2.2.) and is the signature of an Alfvénic structure rather than reconnection. Thus, signatures in the solar wind magnetic field data alone (e.g. current sheet bifurcation) cannot be used to determine the presence or absence of reconnection.

Figures 5 t-y show the magnetic field hodograms of the transitions across the current sheets in the x'-y' and x'-z' planes determined from MVAB (Section 2). The magnetic field structure in the current sheet of the three events are characteristic of rotational discontinuity (RD), with near-circular field rotation in the x'-y' plane, and a finite normal magnetic field $B_{z'}$ (Sonnerup and Ledley 1974). The Alfvénic nature of the magnetic field and velocity variations across these current sheets provides further support for the interpretation that these current sheets are RDs (e.g., Paschmann et al., 2013).

## 6. STATISTICAL PROPERTIES OF RECONNECTION VERSUS ALFVÉNIC EVENTS

To investigate the differences between reconnecting current sheets and Alfvénic structures, we compare the conditions associated with the 21 reconnection exhausts described in section 4 to those associated with the Alfvénic structures discussed in section 5.

Given the very large number of current sheets associated with Alfvénic structures near perihelion (Horbury et al. 2019), we limit the survey to a representative 16-hour interval shown in Figure 5a-c and select a subset of 'clean' current sheets that satisfy the following criteria:

1. The current sheet is well defined, with monotonic or near-monotonic rotation of the magnetic field from one side to the other.
2. The external boundary conditions are relatively stable.
3. The radial magnetic field reverses polarity across the current sheet. We chose this condition mainly because it is well-defined and it limits the number of events. As will be shown in Figure 6a, even with this criterion, we still have a significant number of low magnetic shear events (down to 38º).
4. The current sheet crossing duration is ≥ 1.8s, to allow for at least two plasma measurements inside the current sheet. This is to ensure that the absence of reconnection jet is not due to measurement resolution limitation.

After applying these criteria we were left with 43 events. We analyzed all current sheets following the same procedure as described in Section 4 and examined some key parameters that may control the occurrence of reconnection to see what may distinguish the reconnection from the non-reconnection events. These parameters are: the magnetic shear angle, the β-magnetic shear condition for reconnection, the velocity shears, and the current sheet thickness.

Figure 6 shows the conditions associated with Alfvénic (non-reconnection, left column) versus reconnecting current sheets detected by PSP (middle column). For comparison, we also examined a previously published large data set of 197 reconnection exhausts detected by Wind at 1 AU (right column) (Phan et al. 2010).

## 6.1. Magnetic shear angle

Figure 6 a-c shows no systematic difference in terms of the magnetic shear across the current sheets. The magnetic shear spans a large (mostly in the 20º-150º) range in both reconnection and non-reconnection categories.

## 6.2. Plasma β-magnetic shear condition for the suppression of reconnection

Swisdak et al. (2003, 2010), based on kinetic simulations, predicted that the occurrence of reconnection in a current sheet depends on a combination of the difference in the β on the two sides of the current sheet and the magnetic shear angle, θ, across the current sheet. The underlying physics is related to the diamagnetic drift of the X-line associated with the plasma pressure gradient across the current sheet. Reconnection is deemed to be suppressed if the X-line drift speed along the reconnection outflow (X) direction exceeds the reconnection outflow speed. For a given θ, Swisdak et al. (2010) predicted that reconnection is suppressed if Δβ satisfies the following relation:

$$\Delta\beta > 2 (L/\lambda_i) \tan (\theta/2) \qquad (2)$$

where $L/\lambda_i$ is the width of the plasma pressure gradient layer across the current sheet (near the X-line) in units of the ion skin depth $\lambda_i$. This width is a free parameter but is expected to be comparable to the width of the ion diffusion region which, in turn, is expected to be of the order of $\lambda_i$. According to this prediction, reconnection is allowed for a large range of θ at low Δβ but requires large θ at high Δβ values. Thus, it would be easier for reconnection to occur in low β environments since Δβ would also be small.

Observations of reconnection events in the solar wind at 1 AU (Phan et al. 2010; Gosling and Phan 2013) and at Earth's magnetopause (Phan et al. 2013a; Trenchi et al. 2015) seem to be in agreement with this prediction: For low Δβ, reconnection exhausts were observed for a large range of θ, whereas for large Δβ reconnection occurred only when θ was large. This prediction has also been used to explain the high occurrence rate of reconnection at Mercury's magnetopause (DiBraccio et al. 2013) and the low occurrence rate of reconnection at the equatorial magnetopause at Saturn (Masters et al. 2012).

We now address the Δβ and θ conditions for the PSP events with the goal of determining whether the Δβ and θ conditions in the Alfvénic events may be different from the reconnection events. The analysis uses only proton β because PSP electron temperature data resolution was not sufficiently high for our events. However, based on the Wind spacecraft database of 197 reconnection exhausts at 1 AU (Phan et al., 2010), $\Delta\beta_{electron}$ is somewhat correlated with $\Delta\beta_{proton}$, and comparable in size (not shown). Thus the true Δβ is likely to be larger (but likely less than a factor of two) than $\Delta\beta_{proton}$. This should be kept in mind as one interprets the results in Figures 6d-f.

Figures 6d-f show that, with the exception of one PSP reconnection event (Figure 6e), no other reconnection or non-reconnection events were in the parameter regime where reconnection is predicted to be suppressed (below the relation 2 curve in Figures 6d-f). In fact, the Alfvénic (non-reconnection) events have even smaller $\Delta\beta_{proton}$ on average than the reconnection events, thus safely away from the marginal diamagnetic drift suppression condition. The smaller $\Delta\beta_{proton}$ of the Alfvénic events is due to fact these current sheets were in the lower β region near perihelion. This finding thus reveals that the Δβ−θ condition is not a distinguishing factor between reconnection and non-reconnection PSP events. It further demonstrates that the Δβ−θ reconnection suppression condition is a necessary, but not sufficient condition for reconnection.

## 6.3. Velocity shears

Theories and simulations have predicted that large tangential velocity shears across current sheets exceeding the Alfvén speed can suppress reconnection (e.g., Cowley and Owen 1989; Cassak and Otto 2011). However, as far as we know, the suppression of reconnection by velocity shear has not been demonstrated experimentally. Furthermore, most (if not all) of the theoretical predictions are for the case of zero or small guide field, whereas in the events examined here, there is often a substantial guide field.

We have examined whether the size of the tangential velocity shear may be different for reconnection and non-reconnection events. Figures 6g-i show the distributions of the shear in the velocity component parallel to the reconnecting field, normalized to the shear of the Alfvén velocity based on the reconnecting field, $\Delta V_{pX}/\Delta V_{AX}$. The general finding is that the Alfvénic (non-reconnection) events have much larger velocity shear than the reconnection events, with a median of 0.71, as opposed to 0.24 and 0.12 for the PSP and Wind reconnection events, respectively. This may suggest that local reconnection was suppressed for large (but sub-Alfvénic) velocity shears. We note that the observed velocity shear almost never exceeded unity, i.e. boundary conditions with velocity shears larger than Alfvén velocity shear do not seem to exist in this solar wind (Figure 6g).

While MHD-based theories have predicted (for anti-parallel reconnection with no guide field) the suppression of reconnection to occur when the normalized velocity shear exceeds unity (e.g., Cassak and Otto, 2011), these predictions are related to the reconnection outflow being at the Alfvén speed. However, in observations (including those of PSP described in Section 4.1) as well as in kinetic simulations, the outflow jet speed is generally sub-Alfvénic (e.g., Paschmann et al., 1986; Oieroset et al. 2000; Haggerty et al. 2018). If the reconnection outflow is sub-Alfvénic it seems possible that reconnection could be suppressed at a velocity shear that is sub-Alfvénic.

We also examined the normalized shear of the transverse velocity component $\Delta V_{pY}/\Delta V_{AX}$, and found no significant difference between the reconnection and non-reconnection events, and $\Delta V_{pY}$ was much smaller than $\Delta V_{pX}$ in general.

Another possible explanation for the different velocity shears for the reconnection and Alfvénic events is that current sheets that reconnect are those that were originally TDs (with velocity shears that are unrelated to the local Alfvén speed), whereas the current sheets in Alfvénic structures are RDs, which contain finite normal magnetic fields and Alfvénic velocity shears. There are presumably plasma flows through RDs as well which are usually not measurable. RDs are not known to reconnect.

### *6.4. Current sheet thickness*

As shown in Table 1, PSP reconnection events exhibit a large range of current sheet crossing durations and thicknesses, ranging from 1.6 s (24 ion inertial length $d_i$) to 1150 s (17415 $d_i$), with a median of 14s (170 $d_i$), where the thickness is the product of crossing duration and the average normal velocity $V_{pZ}$ measured at the two edges of the current sheet. For the 43 Alfvénic (non-reconnection) events that we studied, the median current sheet crossing duration and current sheet width were smaller: 7.5 s (90 $d_i$), and the width ranges from 9 $d_i$ to 455 $d_i$.

In collisionless reconnection, it is generally understood that reconnection is triggered only when the thickness of a current sheet is of the order of an ion inertial length or smaller (e.g., Sanny et al. 1994). Thus, it seems that all the (reconnecting and non-reconnecting) current sheets we examined were too thick to reconnect. However, for reconnection events, the relevant current sheet thickness for triggering reconnection is around the X-line. After reconnection is triggered, the exhaust widens with increasing distance from the X-line, and PSP generally crosses the exhaust at some distance downstream of the X-line. On the other hand, the thicker-than-$d_i$-scale Alfvénic current sheets may still be a factor in stabilizing reconnection.

## 7. SUMMARY AND DISCUSSIONS

During the first orbit, PSP encountered reconnection exhausts in ICMEs, in HCS crossings, and in the regular solar wind. Many of the reconnecting current sheets were bifurcated, resembling Petschek's model of reconnection with a pair of slow-shock/RD like structures bounding the exhaust (Petschek 1964; Lin and Lee 1993). All the reconnection exhausts identified so far in Orbit 1 were detected relatively far from the Sun (44.4 - 107.2 $R_s$). About half of the reconnection events had magnetic shear < 90º, and one case had a magnetic shear of 27º, i.e., a guide field of 4. The extreme low-shear current sheets produced plasma jetting as slow as 10 km/s (relative to the external flows). The well-resolved current sheets with clean plasma and field signatures of reconnection clearly demonstrated the capability of PSP to detect reconnection exhausts in the solar wind.

Magnetic reconnection in ICMEs is important as it can cause erosion and changes to the magnetic field structure which may be relevant for understanding their geoeffectiveness at Earth (e.g., Lavraud et al., 2014; Fermo et al., 2014). The PSP measurements presented here show that reconnection within ICMEs can in fact already be operating at 54-55 $R_S$ from the Sun, which means reconnection could work to affect the ICME structure for the majority of its transit from the Sun to 1 AU. Similarly, the detection of well-established reconnection exhausts in the HCS observed inside of 61 $R_s$ indicates that magnetic topology around HCS is already being altered at relatively close distances to the Sun.

Surprisingly, however, we were not able to find reconnection exhausts during the 9 days (2018-11-02/13:15 – 2018-11-11/03:28 UT) around perihelion, even though this period had the highest Alfvén speed and lowest plasma β of Orbit 1. The perihelion period was dominated by Alfvénic structures associated with bursty radial jets which may have originated from a coronal hole (Bale et al. 2019; Badman et al. 2019). Similar Alfvénic structures (or pulses) have previously been detected by the Helios spacecraft further away from the Sun, at ~60 $R_s$ (Horbury et al. 2018), and also at 1 AU (Gosling et al. 2011; Matteini et al. 2014). None of the current sheets associated with these Alfvénic structures that we have examined appear to be undergoing local reconnection at the PSP location.

To identify the possible conditions that may control the onset of local reconnection we compared the plasma and field conditions surrounding current sheets with and without reconnection. The main noticeable difference is the degree of tangential plasma flow shear relative to the Alfvén velocity shear, $\Delta V_{pX}/\Delta V_{AX}$, with Alfvénic events having much larger velocity shear than reconnection events. Our finding may suggest that large, but sub-Alfvénic, velocity shears could suppress reconnection.

An alternative explanation for the absence of reconnection in the Alfvénic current sheets is that these current sheets are RDs (see Section 5). We postulate that, in addition to near-Alfvénic flow shears, the finite normal magnetic field (Buechner and Zelenyi, 1987) and the associated normal plasma flow (at the Alfvén speed based on the normal magnetic field) through the RDs could preclude these current sheets from reconnecting locally. The thicker-than-$d_i$-scale Alfvénic current sheets may also be a factor in stabilizing reconnection. If and how reconnection can be triggered in RDs will need to be investigated in future studies.

It is presently not clear why most of the current sheets near Orbit 1 perihelion are Alfvénic (or RDs) in nature. From just one PSP pass we are not able to determine whether the observed lack of local reconnection in current sheets near perihelion is a general phenomenon or a one-time occurrence associated with the specific solar wind encountered during Orbit 1. In principle, the low-$\beta$, high Alfvén speed environment close to the Sun is expected to be more favorable to the onset of reconnection than the solar wind farther away. Inter-comparison of data from future PSP orbits sampling different types of solar wind will shed further light on this matter.

**Acknowledgments**

We are grateful for the dedicated efforts of the entire Parker Solar Probe team. We thank Yu Lin and Bengt Sonnerup for helpful discussions. The FIELDS experiment was designed and developed under NASA contract NNN06AA01C. J. P. E. and T. S. H. were supported by UK STFC grant ST/S000364/1. TDP was supported by NASA grant 80NSSC18K0157. JFD was supported in part by NSF Grant No. PHY1805829, NASA Grant No. NNX14AC78G. MAS was supported by NASA grant NNX17AI25G.

**Figure Captions**

Figure 1. Overview of Encounter 1 showing (1) current density spikes indicative of the presence of a large number of current sheets throughout the encounter, and (2) the locations of heliopsheric current sheets (HCS) and interplanetary coronal mass ejections (ICME) where reconnection exhausts are seen. (a) Magnetic field magnitude, (b) radial component of the magnetic field, (c) normalized pitch angle energy fluxes of 310 eV (strahl) electrons measured by SWEAP/SPAN, (d) Alfvén speed, and (e) proxy for current density: $|j|= [(\delta B_R/\delta t)^2+(\delta B_T/\delta t)^2+(\delta B_N/\delta t)^2)]^{1/2}/(\mu_0 V_{pR})$, where $V_{pR}$ is the radial component of the proton velocity.

Figure 2. Examples of reconnection exhausts in current sheets associated with an ICME. (a,b) Overview of ICME on 2018-10-31. (c-g) and (h-l) are two extremely low magnetic-shear current sheets inside the ICME displaying reconnection ion jets. (m-q) reconnecting current sheet at the compressed trailing edge of the ICME. (a,b) Magnetic field magnitude and components in RTN coordinates, (c,h,m) magnetic field magnitude, (d,i,n) magnetic field components in XYZ current sheet boundary normal coordinates, (e,j,o) X component of the magnetic field, (f,k,p) X component of the proton bulk velocity measured by SWEAP/SPC, (g,l,q) R component of the proton velocity, and (r) schematic illustrations of reconnection configuration and PSP trajectories across a reconnecting sheet above (+X) and below (-X) the reconnection site. Vertical blue dashed lines mark approximately the two edges of a current sheet where the external conditions have stabilized. The magnetic field and ion velocity values at these two locations are used to compute the magnetic shear angle θ and velocity shear across the current sheet $\Delta V_{pX}$. Panels e and f show how reconnection exhausts are recognized: Correlated changes in $V_{pX}$ and $B_X$ at one edge of a current sheet, and anti-correlated changes in $V_{pX}$ and $B_X$ at the other edge, indicative of Alfvén waves propagating in opposite directions along reconnected field lines away from the reconnection X-line (e.g., Gosling et al. 2005a).

Figure 3. Examples of reconnection exhausts at heliospheric current sheets. (a) Radial component of the magnetic field, (b) normalized pitch angle energy fluxes of 310 eV (strahl) electrons. The parameters in panels c-v are the same as those in Figure 2. $V_{pR}$ was positively enhanced in all 4 current sheets, indicating that the reconnection X-line was sunward of PSP in these events.

Figure 4. Examples of reconnection exhausts in the regular solar wind. (a) Radial component of the magnetic field, (b) proxy for current density (see Figure 1 caption). The parameters in panels c-l are similar to those in Figure 2. The event in panels h-l illustrates the high quality of 0.22s-resolution SWEAP/SPC proton velocity measurements, which fully resolved the proton reconnection jet in the 5.8s current sheet crossing.

Figure 5. Current sheets close to perihelion are mostly associated with Alfvénic structures rather than reconnection. (a,b) Radial component of the magnetic field and ion bulk velocity, (c) overlaid radial proton velocity and radial Alfvén velocity, and (d) zoomed in of Panel c. The parameters in panels e-s are similar to those in Figure 2. (t-y) Magnetic field hodograms in the x'-y' and x'-z' planes determined from MVAB (Section 2). The normal magnetic field errors were estimated using the method by Sonnerup and Scheible (1998). Panels g and h show how to recognize Alfvénic structures as opposed to reconnection: $V_{pX}$ and $B_X$ variations are positively correlated at both edges of the current sheet. This single (positive) correlation between $V_{pR}$ and $V_{AR}$ persisted throughout the 16-hour interval, indicating that $V_{pR}$ enhancements (from a baseline of ~ 300 km/s) are Alfvénic structures, not reconnection jets.

Figure 6. Statistical properties of 43 PSP non-reconnecting (Alfvénic) current sheets associated with $B_R$ polarity reversals in the 16-hour interval of Figures 6a-c (left column), 21 reconnecting current sheets found in PSP Encounter 1 (middle column), and 197 Wind reconnecting current sheets from Phan et al. (2010) (right column). (a-c) Magnetic shear angle θ, (d-f) difference in proton β across the current sheet versus θ, (g-i) velocity shear across the current sheets in the X direction, normalized to Alfvén velocity shear in same direction, and (j-l) velocity shear across the current sheets in the out-of-plane Y direction, normalized to Alfvén velocity shear in the X direction. The solid curve in Panels d-f is the diamagnetic drift suppression marginal condition Δβ=2 tan(θ/2), below which reconnection is predicted to be suppressed (Swisdak et al. 2003; 2010). All the Alfvénic events are in the area above the curve (Panel d), thus the lack of reconnection in Alfvénic current sheets is not due to the diamagnetic drift suppression effect. The main noted difference between reconnection (Panels h,i) and non-reconnection (Panel g) events is the higher velocity shears (in X) across Alfvénic (non-reconnecting) current sheets.

**Table 1**
List of reconnection events

| Event num | Start time[a] (UT) | Dur[b] (s) | Width (km) | Width ($d_i$) | Distance to X-line ($R_S$) | Magnetic shear (degrees) | Guide field (nT) | $\Delta V_{pX}/\Delta V_{AX}$ | $\Delta V_{pY}/\Delta V_{AX}$ | $V_R$ in reconnection[c] | Distance to Sun ($R_S$) | Context |
|---|---|---|---|---|---|---|---|---|---|---|---|---|
| 1 | 2018-10-27/04:42:00 | 51.9 | 4034 | 198 | 0.029 | 133 | 0.44 | 0.10 | 0.079 | ↑ | 74.49 | SW |
| 2 | 2018-10-29/19:07:30 | 38.4 | 7729 | 422 | 0.056 | 120 | 0.58 | 0.30 | 0.062 | ? | 62.01 | SW |
| 3 | 2018-10-31/03:38:03 | 1.60 | 405 | 24 | 0.0029 | 49 | 2.2 | 0.0043 | 0.19 | ? | 55.42 | ICME |
| 4 | 2018-10-31/06:43:20 | 5.71 | 1419 | 88 | 0.010 | 35 | 3.2 | 0.18 | 0.043 | ↑ | 54.80 | ICME |
| 5 | 2018-10-31/07:18:50 | 6.29 | 1718 | 97 | 0.012 | 27 | 4.2 | 0.21 | 0.045 | ↓ | 54.68 | ICME |
| 6 | 2018-10-31/08:31:03 | 7.61 | 2143 | 139 | 0.015 | 89 | 1.0 | 0.38 | 0.33 | ↓ | 54.44 | ICME |
| 7 | 2018-10-31/12:14:30 | 32.4 | 3481 | 183 | 0.025 | 97 | 0.89 | 0.56 | 0.35 | ↑ | 53.70 | SW |
| 8 | 2018-10-31/14:54:32 | 5.60 | 1128 | 70 | 0.0081 | 128 | 0.48 | 0.34 | 0.23 | ↓ | 53.17 | SW |
| 9 | 2018-10-31/14:57:10 | 4.11 | 969 | 51 | 0.0070 | 74 | 1.3 | 0.39 | 0.17 | ↓ | 53.16 | SW |
| 10 | 2018-11-01/08:53:33 | 2.01 | 320 | 20 | 0.0023 | 88 | 1.0 | 0.24 | 0.012 | ↓ | 49.64 | SW |
| 11 | 2018-11-01/23:23:07 | 3.85 | 473 | 31 | 0.0034 | 53 | 2.0 | 0.047 | 0.28 | ↑ | 46.91 | SW |
| 12 | 2018-11-01/23:25:03 | 27.8 | 2596 | 174 | 0.019 | 60 | 1.7 | 0.13 | 0.34 | ↓ | 46.90 | SW |
| 13 | 2018-11-02/12:42:45 | 5.80 | 1626 | 104 | 0.012 | 55 | 1.9 | 0.064 | 0.13 | ↓ | 44.52 | SW |
| 14 | 2018-11-02/13:15:00 | 7.33 | 1584 | 98 | 0.011 | 83 | 1.1 | 0.26 | 0.063 | ↓ | 44.42 | SW |
| 15 | 2018-11-11/23:17:00 | 65.7 | 24642 | 1480 | 0.18 | 101 | 0.82 | 0.067 | 0.51 | ↑ | 54.62 | ICME |
| 16 | 2018-11-13/07:11:00 | 28.4 | 3139 | 170 | 0.023 | 129 | 0.48 | 0.36 | 0.11 | ↑ | 61.07 | HCS |
| 17 | 2018-11-13/10:17:00 | 42.6 | 12852 | 800 | 0.092 | 71 | 1.4 | 0.48 | 0.13 | ↓ | 61.71 | HCS |
| 18 | 2018-11-13/16:15:00 | 1150 | 287940 | 17415 | 2.1 | 168 | 0.11 | 0.41 | 0.0039 | ↑ | 62.92 | HCS |
| 19 | 2018-11-13/23:07:00 | 14.1 | 4466 | 296 | 0.032 | 133 | 0.43 | 0.29 | 0.049 | ↑ | 64.30 | HCS |
| 20 | 2018-11-14/13:20:00 | 192.0 | 30687 | 1694 | 0.22 | 108 | 0.73 | 0.087 | 0.28 | ↑ | 67.18 | HCS |
| 21 | 2018-11-23/18:26:00 | 650.0 | 171767 | 5965 | 1.2 | 168 | 0.10 | 0.072 | 0.17 | ↑ | 107.20 | HCS |

[a] left edge of current sheet
[b] current sheet crossing duration
[c] radial velocity in reconnection exhaust: Enhanced (↑), reduced (↓), or ambiguous (?).

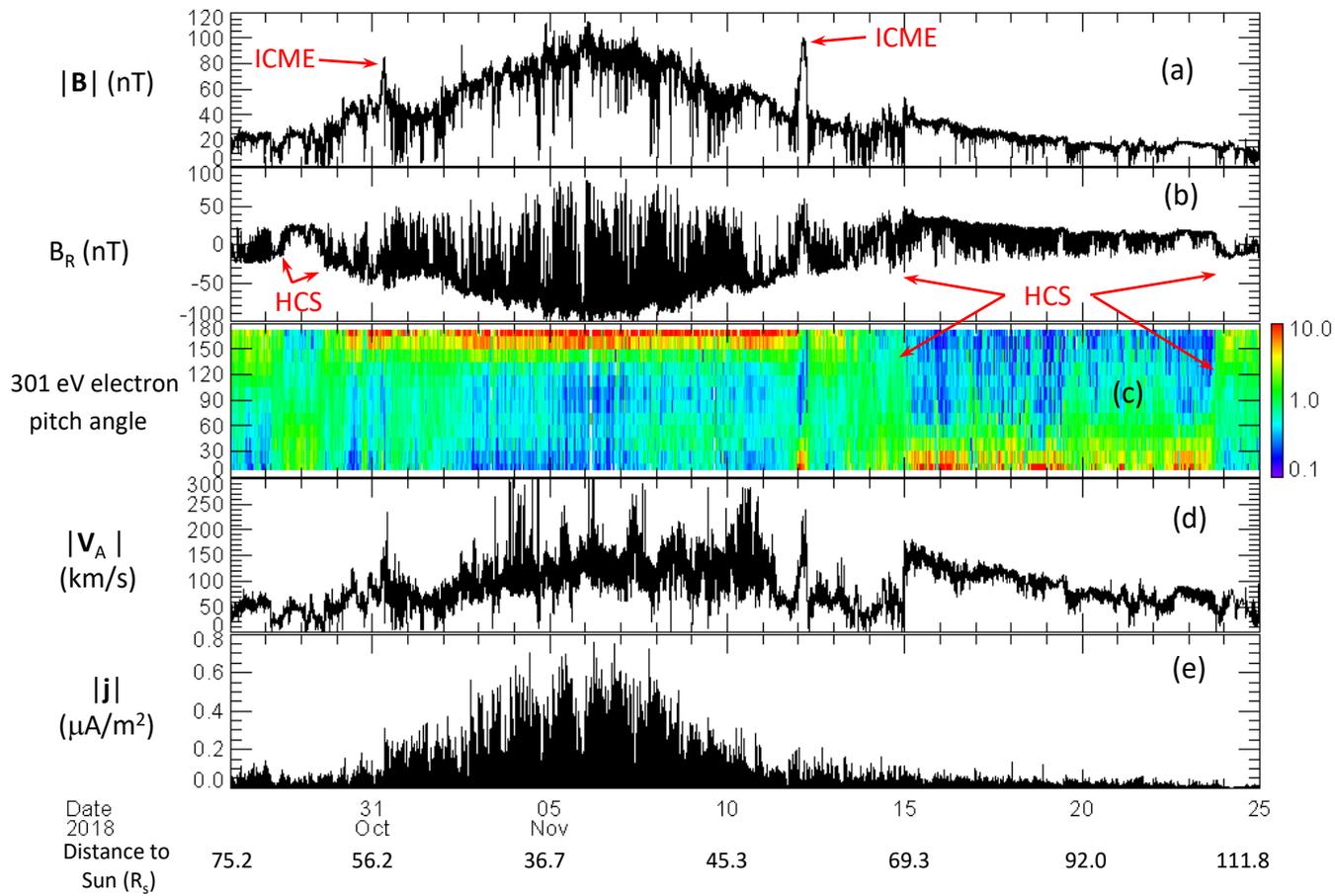

# Reconnection Exhausts in Current Sheets inside Oct 31 ICME

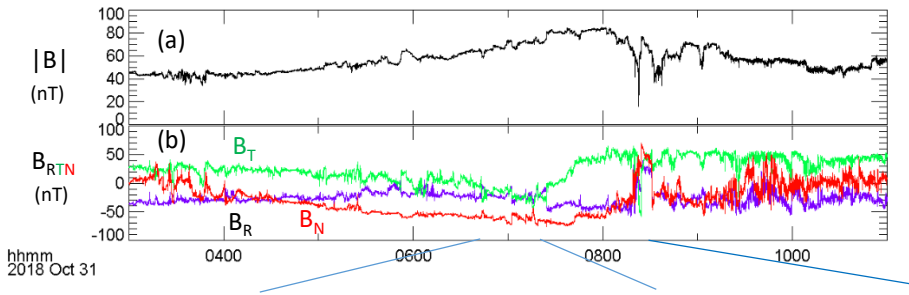
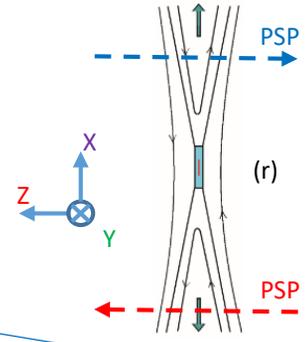
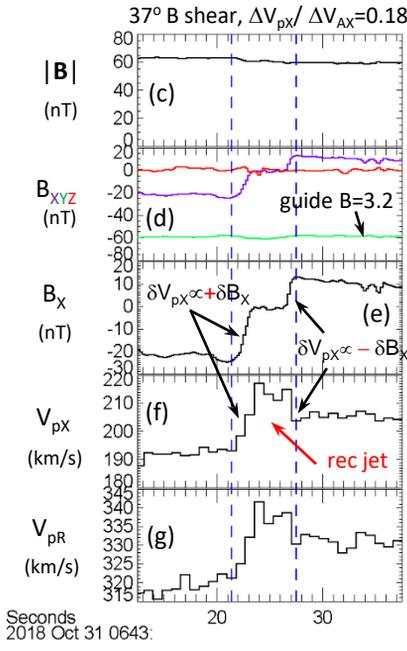
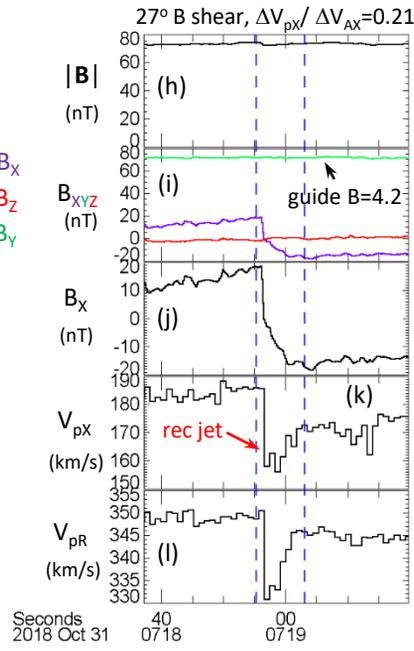
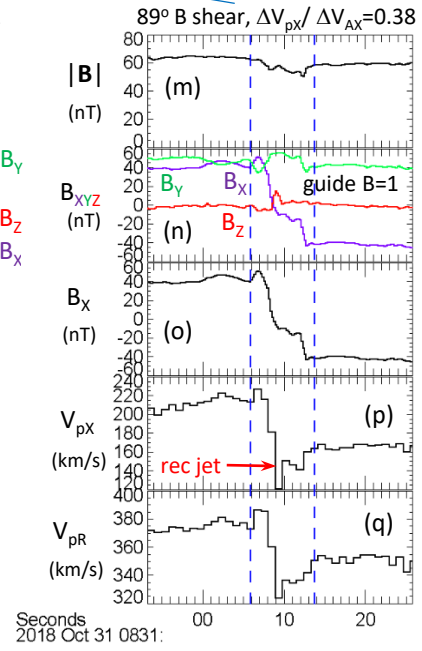

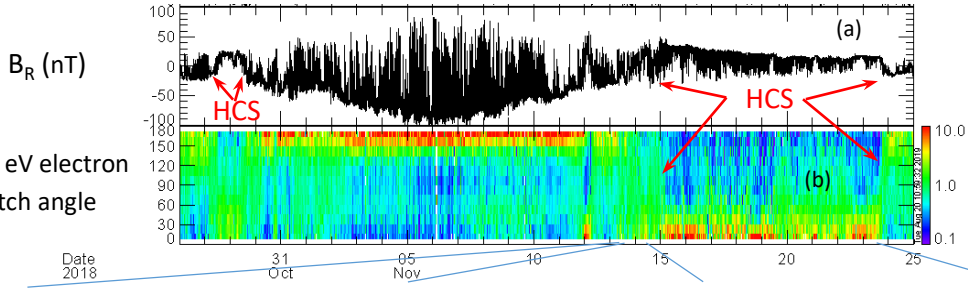
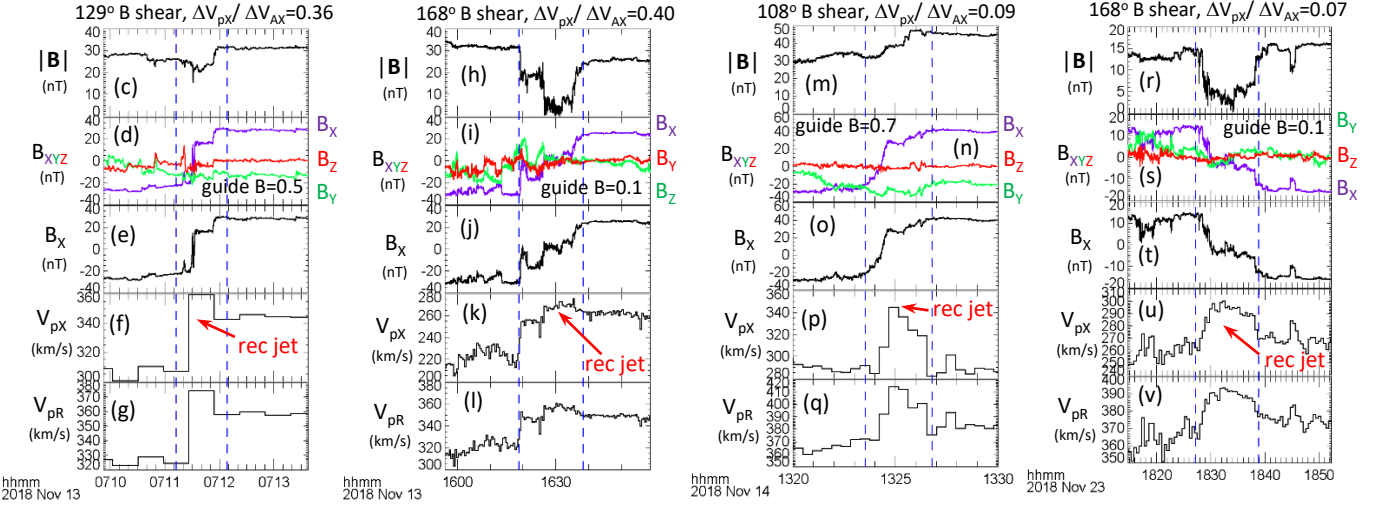

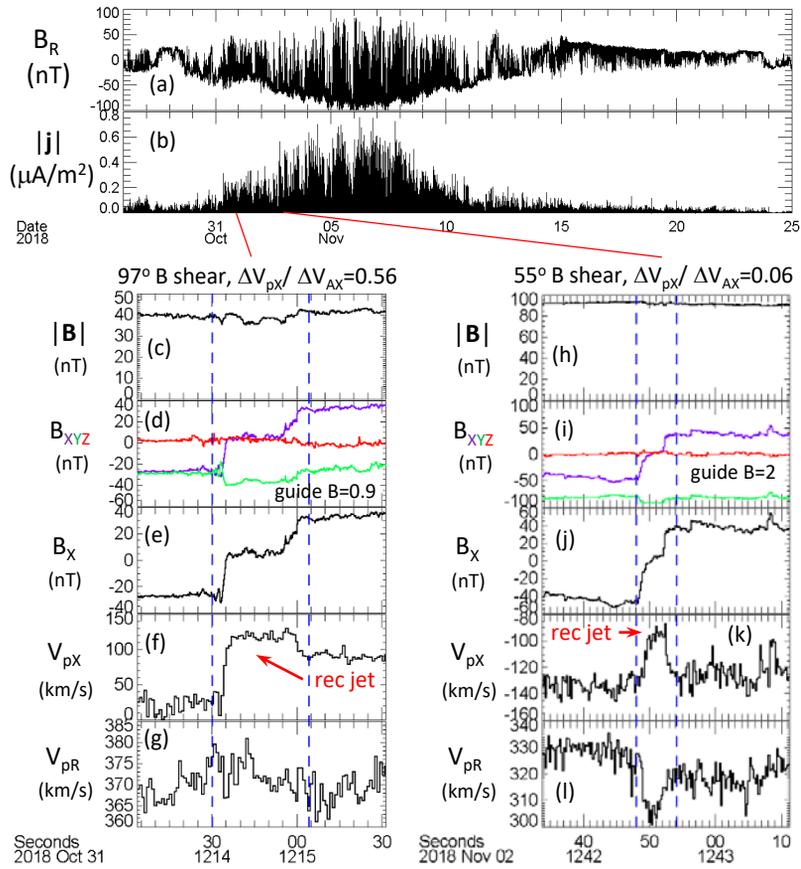

# (Non-Reconnection) Alfvénic Structures near Perihelion

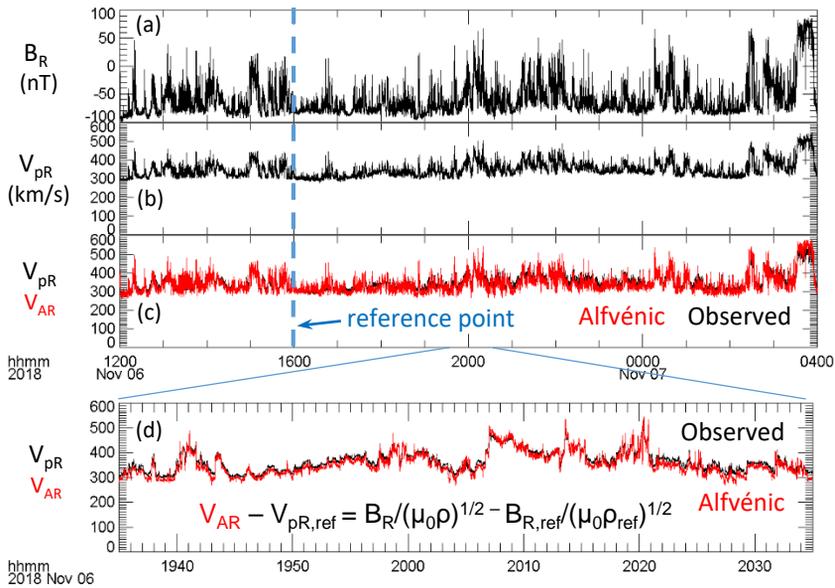

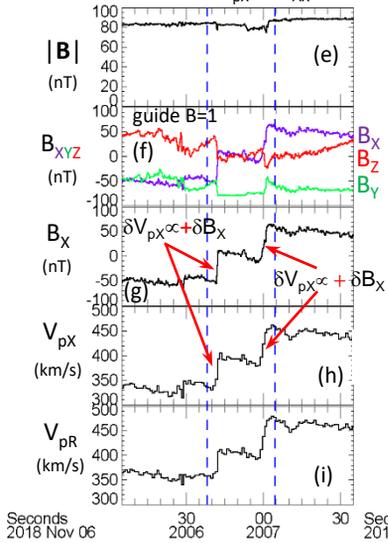
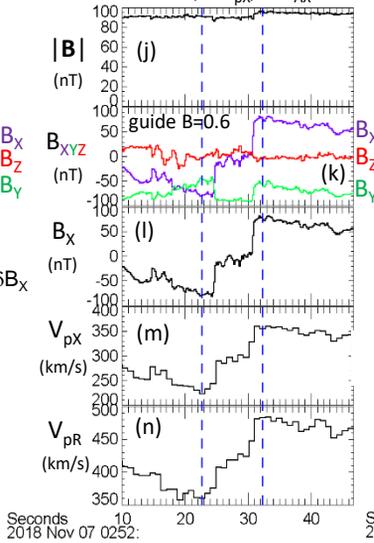
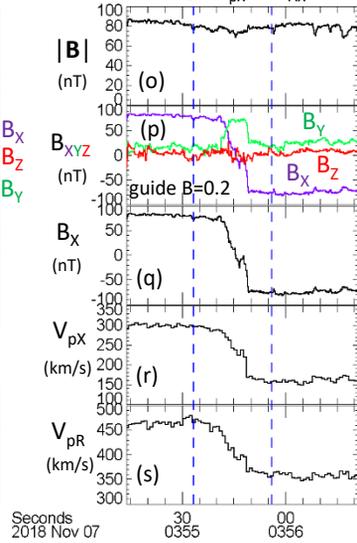

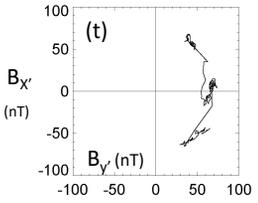
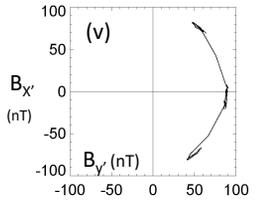
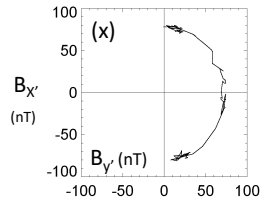

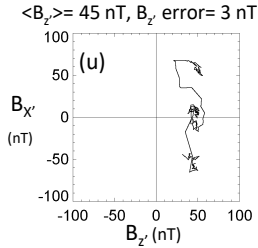
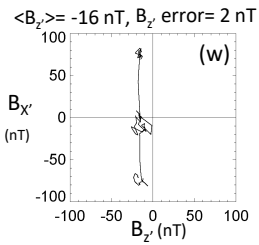
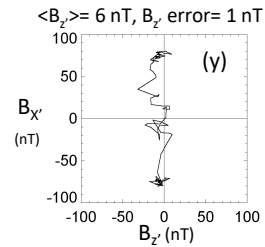

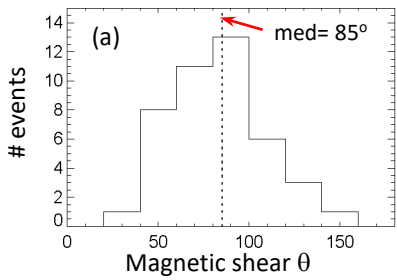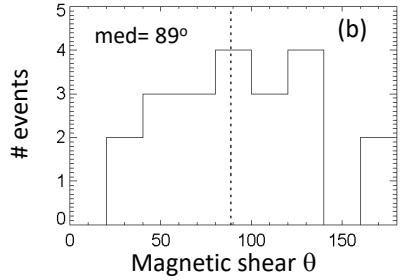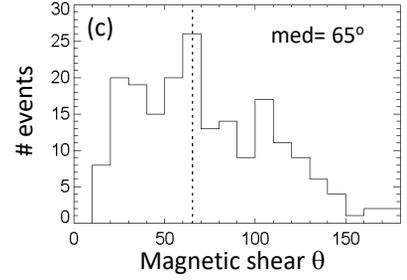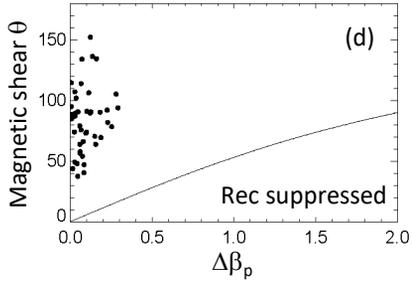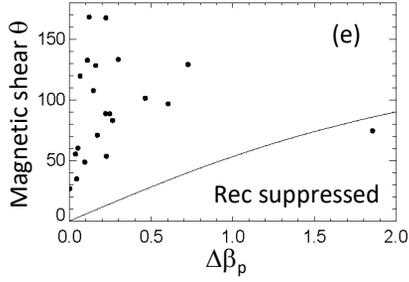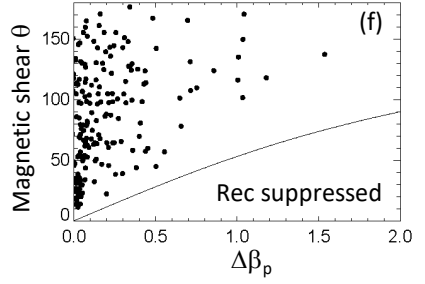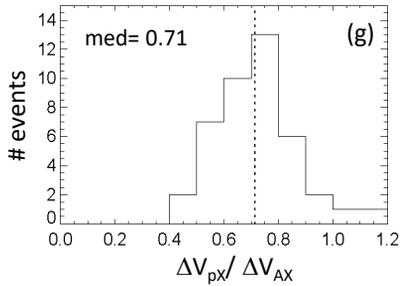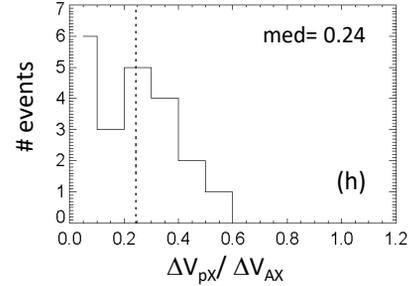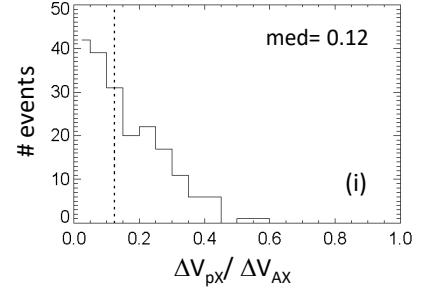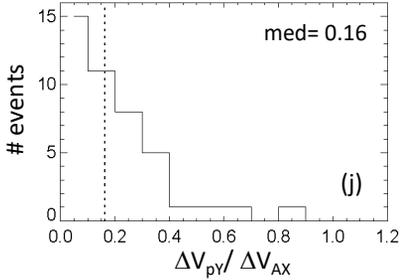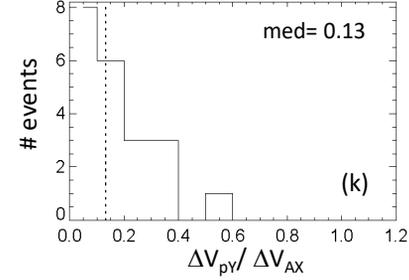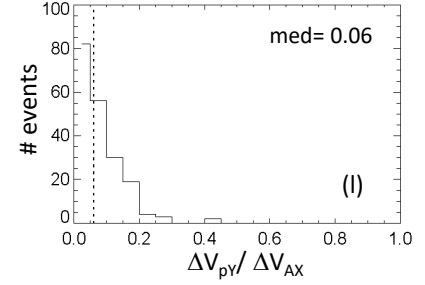